\documentclass[11pt]{wlscirep}
\usepackage{graphicx}
\usepackage{hyperref}
\usepackage[T1]{fontenc}

\title{Unconventional pairing in few-fermion systems at finite temperature}
   
\author[1,2,*]{Daniel Pecak}
\author[1,3]{Tomasz Sowi\'{n}ski}
\affil[1]{Institute of Physics, Polish Academy of Sciences, Aleja Lotnik\'{o}w 32/46, PL-02668, Warsaw, Poland}
\affil[2]{Faculty of Physics, Warsaw University of Technology, Ulica Koszykowa 75, PL-00662 Warsaw, Poland}
\affil[3]{\mbox{Instituto Universitario de Matem\'atica Pura y Aplicada, Universitat Polit\`ecnica de Val\`encia, ES-46022 Val\`encia, Spain}}
\affil[*]{daniel.pecak@ifpan.edu.pl}
 
\begin{abstract}
Attractively interacting two-component mixtures of fermionic particles confined in a one-dimensional harmonic trap are investigated. Properties of balanced and imbalanced systems are systematically explored with the exact diagonalization approach, focusing on the finite-temperature effects. Using single- and two-particle density distributions, specific non-classical pairing correlations are analyzed in terms of the noise correlations -- quantity directly accessible in state-of-the-art experiments with ultra-cold atoms. It is shown that along with increasing temperature, any imbalanced system hosting Fulde-Ferrel-Larkin-Ovchinnikov pairs crossovers to a standard Bardeen-Cooper-Schrieffer one characterized by zero net momentum of resulting pairs. By performing calculations for systems with different imbalances, the approximate boundary between the two phases on a phase diagram is determined.  
\end{abstract}
\begin{document}

\flushbottom
\maketitle
\section{Introduction}
The appearance of non-classical correlations collectively induced by attractive interactions between fermions is one of the most spectacular macroscopic manifestations of quantum mechanics. As justified by Bardeen, Cooper, and Schrieffer in \cite{1975BardeenPhysRev} (BCS), it is directly responsible for the remarkable phenomenon of superconductivity -- a flow of an electric current without any resistance -- discovered by Onnes over a hundred years ago~\cite{1911Onnes}. In the case of homogeneous bulk systems, an existence of pairing induced by attractions is explained on fundamental grounds by showing that for any non-vanishing attractions the system is forced in a nonperturbative manner to rebrand its ground state and minimize energy by forming pairs (called Cooper pairs) composed by opposite-spin fermions having exactly opposite momenta~\cite{1956CooperPhysRev,1971FetterBook}. Later it was argued that this argumentation is rigorously valid only for perfectly balanced systems, {\it i.e.}, when both components have exactly the same Fermi spheres. In a more general scenario, when the system is no longer symmetric with respect to the exchange of components, pairing correlations are still favored but resulting pairs may have non-vanishing center-of-mass momentum, as explained independently by Fulde and  Ferrel~\cite{1964FuldePhysRev} and Larkin and Ovchinnikov~\cite{1965LarkinJETP} (FFLO). This observation triggered many investigations aimed to understand different properties of such systems (see for example \cite{1966GruenbergPRL,1969TakadaProgThPhys,1990RMPMicnas,1994BurkhardtAnnPhys,2004CasalbuoniRMP,2007MatsudaJPSJ,2012PtokJSNM,2017PtokPRA,2018PtokJSupNovMag,2019MagierskiPRA,2020ZdybelPRR,2021MagierskiPRA} and citations within) and in consequence to unquestionably confirm that such pairs do exist. Unfortunately, indisputable experimental evidence is still lacking. In three-dimensional cases, one of the recognized obstacles is the fact that the final FFLO signal is much less noticeable since any non-zero value of finite net momentum has no unique direction in space. In consequence, the condensation of pairs is spread among different two-particle orbitals and the FFLO phase is stable in a quite small region of the phase diagram~\cite{2007ParishNaturePhys}. It is argued that systems with reduced dimensionality (particularly one-dimensional systems) may serve as a promising bypassing platforms for FFLO detection~\cite{2001YangPRB,2007OrsoPRL,2007FeiguinPRB,2008BatrouniPRL,2008RizziPRB,2010ZapataPRL,2010LiaoNature,2020RammelSciPostPhys}. For recent reviews see~\cite{2018KinnunenRPP,2020DobrzynieckiAdvQTech}. The universal nature of such correlations in different number of spatial dimensions can be seen through the variety of physical systems that are studied. From nuclear matter and neutron stars \cite{alford2001colour,cirigliano2011low,2019SedrakianEPJA}, through organic superconductors \cite{FFLOOrganic1,FFLOOrganic2,FFLOOrganic3} and heavy-fermion systems in solid state physics \cite{FFLOHeavy1,FFLOHeavy2,2007MatsudaJPSJ,FFLOHeavy4}, up to ultracold gases \cite{2018KinnunenRPP}.

Significant experimental progress on ultracold atomic systems has opened many interesting and non-obvious paths for exploration. One of them is related to the ability of very precise control of the number of particles in a system. Pioneering experiments of this type~\cite{2008CheinetPRL,2013ZurnPRL,2011SerwaneScience} have opened an alternative route to study quantum many-body systems from the side of mesoscopic phenomena~\cite{2012BlumeRPP,2019SowinskiRPP}. Amidst a flurry of different ideas related to such small quantum systems, recently it has been theoretically spotted that small mixtures of attractively interacting fermions may host correlations very similar to pairing known for bulk systems~\cite{1997MatveevPRL,2015SowinskiEPL,2015DamicoPRA,2016HofmannPRA,2020PecakPRR, 2020LydzbaPRA, 2021DobrzynieckiPRR}. Depending on inter-component imbalance, they may serve as suitable platforms for observing unconventional pairing mechanisms~\cite{2021SowinskiEPLReview}. Very recently, these theoretical considerations have taken on new importance since innovative methods for detecting and measuring inter-particle correlations in such systems have been just developed~\cite{2016CheukScience,2018BergschneiderPRA,2019BergschneiderNaturePhys}. The majority of previous works have mainly focused on different ground-state properties or some specific dynamical scenarios of such systems with repulsive interactions. In this work, we aim to go beyond these schemes and to extend recent discussion on pairing (the conventional BCS as well as the unconventional FFLO) to the situation when the system is prepared in a mixed state of a given temperature. In this way, we want to establish another theoretical link between few-body problems and finite-temperature results obtained recently for a large number of particles. Particularly, we want to make a correspondence with previously obtained Monte Carlo predictions and solutions based on the Bogoliubov-de Gennes method for harmonically trapped mixtures of fermions~\cite{2008CasulaPRA,2010WolakPRA,machida2006generic}. It is worth noticing that on the repulsive branch of interactions some theoretical analyses of the impact of finite temperatures were presented already~\cite{2013SowinskiPRA,2018YaoPRL,2018PlodzienPRA,2020CapuzziPRA}. Although finite-temperature effects for attractively interacting few-fermion systems were not deeply studied, they can be important from the experimental perspective, since BCS-like correlation has been already measured \cite{holten2021observation}.

\section{The system studied}\label{sec:system}
We consider an effectively one-dimensional mixture of two fermionic components $\sigma\in\{\uparrow, \downarrow\}$ consisting of $N_\uparrow$ and $N_\downarrow$ particles of the same mass $m$, and trapped in an external harmonic oscillator potential of frequency $\omega$. We assume that on a time scale of considered experiments there are no relevant physical channels causing spin flips, thus the numbers $N_\uparrow$ and $N_\downarrow$ are independently conserved quantities. In the case of experiments with ultra-cold alkaline atoms, this assumption is well-justified. In the second-quantization formalism the Hamiltonian of the system reads:
\begin{equation}\label{eq:hamiltonian}
\hat{\cal H} = \int\!\!dx\sum_{\sigma} \hat\Psi^\dag_\sigma(x) \left( -\frac{\hslash^2}{2m} \frac{d^2}{dx^2} + \frac{m\omega^2}{2} x^2 \right) \hat\Psi_\sigma(x)
 + g\, \sqrt{\frac{\hslash^3\omega}{m}}\int\!\!dx\, \hat\Psi_\downarrow^\dag(x)\hat\Psi_\uparrow^\dag(x)\hat\Psi_\uparrow(x) \hat\Psi_\downarrow(x),
\end{equation}
where the fermionic field operator $\hat\Psi_\sigma(x)$ annihilates a particle from a component $\sigma$ at position $x$. It obeys standard anti-commutation relations $\{\hat\Psi_\sigma(x),\hat\Psi^\dagger_{\sigma'}(x')\}=\delta_{\sigma\sigma'}\delta(x-x')$ and $\{\hat\Psi_\sigma(x),\hat\Psi_{\sigma'}(x')\}=0$. The dimensionless parameter $g$ is related to the one-dimensional interaction strength. In the effective one-dimensional model studied, it can be obtained from the three-dimensional counterpart, by assuming that transverse motion is completely frozen (for example by very strong confinement) and thus it can be integrated out~\cite{1998OlshaniiPRL}. It means that the value of $g$ depends on the transverse width of the system. This gives one of possible ways to tune its value. Another possibility is granted thank to internal electronic structures of particles, which, although completely irrelevant for spatial dynamics, give a tool for controlling inter-particle scattering via the Feshbach resonances~\cite{2013ZurnPRL,2010ChinRevModPhys}. Since the Hamiltonian does commute with number operators $\hat{N}_\sigma=\int dx\,\hat{\Psi}^\dagger(x)\hat{\Psi}(x)$ we analyze its properties in subspaces of fixed $N_\uparrow$ and $N_\downarrow$. To get better correspondence with standard description in the large system limit, we introduce an intensive quantity characterizing imbalance between components, {\it i.e.}, the global polarization 
\begin{equation} \label{polarization}
\mathrm{P} = \frac{\Delta N}{N}=\frac{N_\uparrow - N_\downarrow}{N_\uparrow+N_\downarrow}.
\end{equation}
Note, that the Hamiltonian \eqref{eq:hamiltonian} is fully symmetric under exchange of components. It means that systems with opposite global polarisations ($\mathrm{P}$ and $-\mathrm{P}$) have exactly the same properties provided that labels $\uparrow$ and $\downarrow$ are adequately exchanged. Therefore, without losing generality, we can limit ourselves to cases with $\mathrm{P}\geq 0$.

\section{Many-body spectrum}\label{sec:spectrum}
To get a better understanding of the properties of the system at finite temperature, let us first discuss spectral properties of the many-body Hamiltonian \eqref{eq:hamiltonian} on the attractive branch of interactions. In contrast to the repulsive one\cite{2013SowinskiPRA,2013GharashiPRL}, this has not been extensively studied in the literature. It is quite natural that in the case studied the spectrum is not bounded from below when attractive $g$ is increasing. In fact, the model \eqref{eq:hamiltonian} is not appropriate for sufficiently large attractions since it completely neglects the possible formation of molecules. However, for finite and reasonable $g$, the model appropriately captures the system's properties \cite{2015DamicoPRA,2015SowinskiEPL} and is experimentally relevant \cite{2013ZurnPRL}. 
 
\begin{figure}[t!]
\centering
\includegraphics[width=\linewidth]{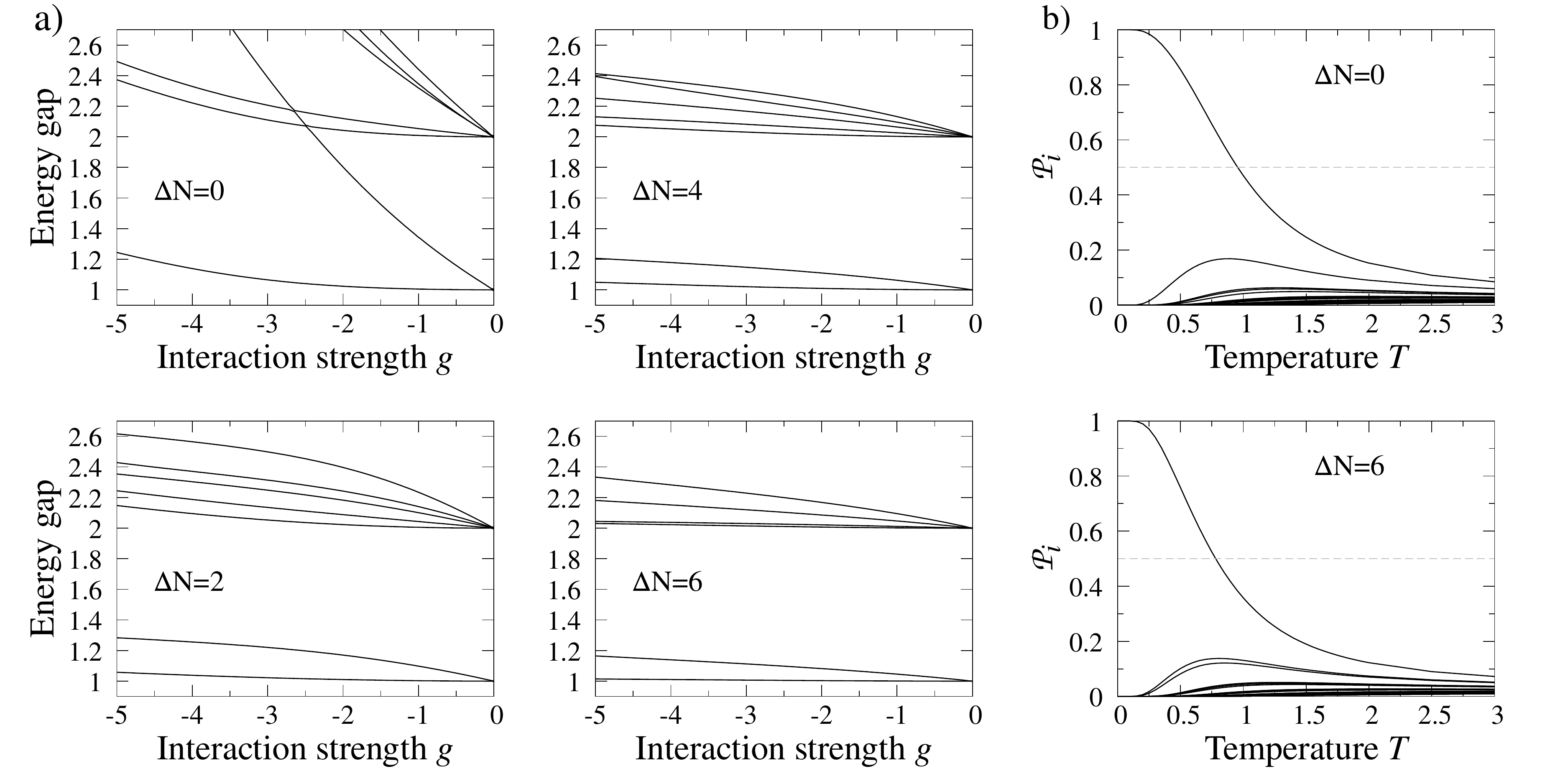}
\caption{(a) The energy gaps between the many-body ground state and excited states for systems with $N=8$ particles and different particle imbalances $\Delta N$ (polarization $\mathrm{P}=0,0.25,0.5,0.75$, respectively). Independently on the case, two of the lowest excited states are always degenerate at $g=0$ and the degeneracy is lifted for finite attraction. Note, however, that the lift strongly depends on the polarization. (b) the Boltzmann probabilities ${\cal P}_i$ as functions of rescaled temperature for two extreme imbalances ($\mathrm{P}=0$ and $P=0.75$, respectively) and interaction strength $g=-3$. Note that for sufficiently low but finite temperatures (due to the finite gap in the spectrum) the system is robust and rather remains in its many-body ground state. All energies and temperatures are expressed in natural units of the problem, {\it i.e.}, $\hslash\omega$ and $\hslash\omega/k_B$, respectively.
\label{Fig1}}
\end{figure}

To present fundamental features of the many-body spectrum of the system studied, we take as an example the system containing $N=8$ particles with different imbalances $\Delta N$ and we perform numerically exact diagonalization of its many-body Hamiltonian \eqref{eq:hamiltonian} to obtain its eigenstates $|{\mathtt i}\rangle$ and corresponding eigenenergies ${\cal E}_i$ (see Appendix~\ref{App1} for numerical details). Then we plot the lowest energy gaps ${\cal E}_i-{\cal E}_0$ as functions of interaction strength $g$ (see Fig.~\ref{Fig1}a). It is clear that for any attraction the many-body ground state is rather isolated. Indeed, for any number of particles and attractive interactions, the gap to the first excited state is never smaller than the energy of a single excitation in the harmonic confinement. It means that any pairing correlations present in the ground-state of the system probably remain stable against small thermal excitations. It is also clear that along with increasing attractions the degeneracy of the first excited state (existing for vanishing interactions due to the fundamental independence of components) is lifted. Note, however, that only for perfectly balanced systems ($\Delta N=0$) one of the first excited states rapidly gains energy and very quickly become unimportant for low-energy physics. That suggests that the thermal resistivity of the system could be strongly dependent on the imbalance, thus may be of fundamental importance for experimental discrimination between standard BCS and FFLO pairing hosted respectively by balanced and imbalanced systems. 

In the following, we assume that the system studied is prepared in a stationary state of a given absolute temperature $T$. We do not settle whether thermalization occurred as a result of interactions with an external thermostat or as a consequence of self-thermalization. The latter may be challenging due to the intrinsic properties of the system -- its lowered dimensionality and mesoscopic character. Nevertheless, we anticipate that the state of the system can be well-described with a density matrix of the Boltzmann form:
\begin{equation} \label{ThState}
 \hat{\rho}_T= \sum_i {\cal P}_i |\mathtt{i}\rangle\langle \mathtt{i}|,
\end{equation}
where ${\cal P}_i={\cal Z}^{-1}\mathrm{e}^{- E_i/k_B T}$ are the Boltzmann distribution probabilities and ${\cal Z} = \sum_i \mathrm{e}^{-E_i/k_B T}$ is the corresponding partition function ($k_{\mathrm{B}}$ is the Boltzmann's constant). Of course, in the limit of vanishing temperature, $T\rightarrow 0$, the state of the system is represented by its many-body ground state $|\mathtt{0}\rangle$. However, even for finite temperatures, due to the mentioned finite gaps to the excited states, the system is quite robust and may remain in the ground state. This property is clearly visible when partitions ${\cal P}_i$ are plotted as functions of temperature (Fig.~\ref{Fig1}b). Since it will be crucial for the discussion of experimental attainability of pairing phases for realistic systems let us relate this observation to the experiments in Selim Jochim's group. From Fig.~\ref{Fig1}b it is clear that the lowest excited states start to contribute at a temperature around $T\approx 0.5 \hslash\omega/k_B$. Taking experimentally typical frequency of the trap $\omega= 2\pi\times 1.488$~kHz and lithium $^6$Li atom mass $m=1.15\times10^{-26}$~kg it corresponds to temperature $35.7$~nK. Thus, we see that the temperature effects definitely have to be taken into account in the modeling when predicting the system properties.

\section{Two-body correlation and pairing}
There are many ambiguities in translating concepts suitable for many-body description into the few-body regime. Mostly, they originate from the fact that in a large number of particles limit one does not care much about their precise counting and therefore performs analysis in the grand canonical ensemble approach. This gives a rise to define mean-field order parameters (pairing fields) as expectation values of appropriately defined annihilation operators~\cite{2006SedrakianBook}. Unfortunately, for small systems we rather focus on properties of the system when the number of particles is well-defined, thus we work in the canonical ensemble framework. Therefore, in the case of mesoscopic systems studied here, instead of considering order parameters, which in fact are not directly measurable, we focus on quantities straightly capturable in experiments that may quantify non-classical pairing correlations in the system. When two-component systems are considered, then the natural quantity describing two-particle correlations collectively induced between particles is the reduced two-particle density matrix. However, from the experimental point of view, measuring its off-diagonal elements (encoding two-body coherence) is not easy if possible at all. Therefore, we focus only on its diagonal part $\mathrm{Tr}\left[\hat{\rho}_T\,\hat{n}_\uparrow(p_1) \hat{n}_\downarrow(p_2)\right]$, {\it i.e.}, the two-particle momentum distribution in the thermal state $\hat\rho_T$. The density operator in the momentum domain $\hat{n}_\sigma(p)$ is defined straightforwardly as $\hat{n}_\sigma(p)=\hat{\Psi}^\dagger_\sigma(p)\hat{\Psi}_\sigma(p)$, where $\Psi_\sigma(p)=\int dx\,\hat{\Psi}_\sigma(x)\mathrm{exp}(-ipx/\hslash)$. 

It is clear that the two-particle momentum distribution defined above encodes not only non-classical correlations forced by interactions but also accidental correlations originating in single-particle distributions \cite{2004AltmanPRA}. Therefore, to filter out the latter from the description, we go along previous experience \cite{2004AltmanPRA,2009MatheyPRA,2008MatheyPRL,2005FollingNature} and we define the so-called noise correlation encoding pure two-body correlations between particles
\begin{equation} \label{ThNoiseCorr}
{\cal G}_T(p_1,p_2)=\langle\hat{n}_\uparrow(p_1) \hat{n}_\downarrow(p_2)\rangle_T-\langle\hat{n}_\uparrow(p_1)\rangle_T\langle\hat{n}_\downarrow(p_2)\rangle_T,
\end{equation}
where we use a short notation $\langle \bullet \rangle_T \equiv \mathrm{Tr}\left[\hat{\rho}_T \,\bullet \,\right]$ for averaging over the many-body mixed state $\hat\rho_T$ of the system of given temperature $T$. Let us point here, that in the case of ground-state properties, the noise correlation concept has been extremely useful to study two-body correlations in general~\cite{2017BrandtPRA}, but the pairing in particular~\cite{2008LuscherPRA,2019PecakPRA, 2020RammelSciPostPhys,2020PecakPRR,2021DobrzynieckiPRR}. It turned out that, at least for the ground-state properties, this measurable quantity can be quite easily adopted as a direct indicator of pairing correlations signaling not only standard Cooper pairs but also unconventional pairs with non-vanishing center-of-mass momentum (FFLO)~\cite{2020PecakPRR}. In the following, we extend this analysis to thermal mixed states of the system. 

\begin{figure}[t!]
\centering
\includegraphics[width=\linewidth]{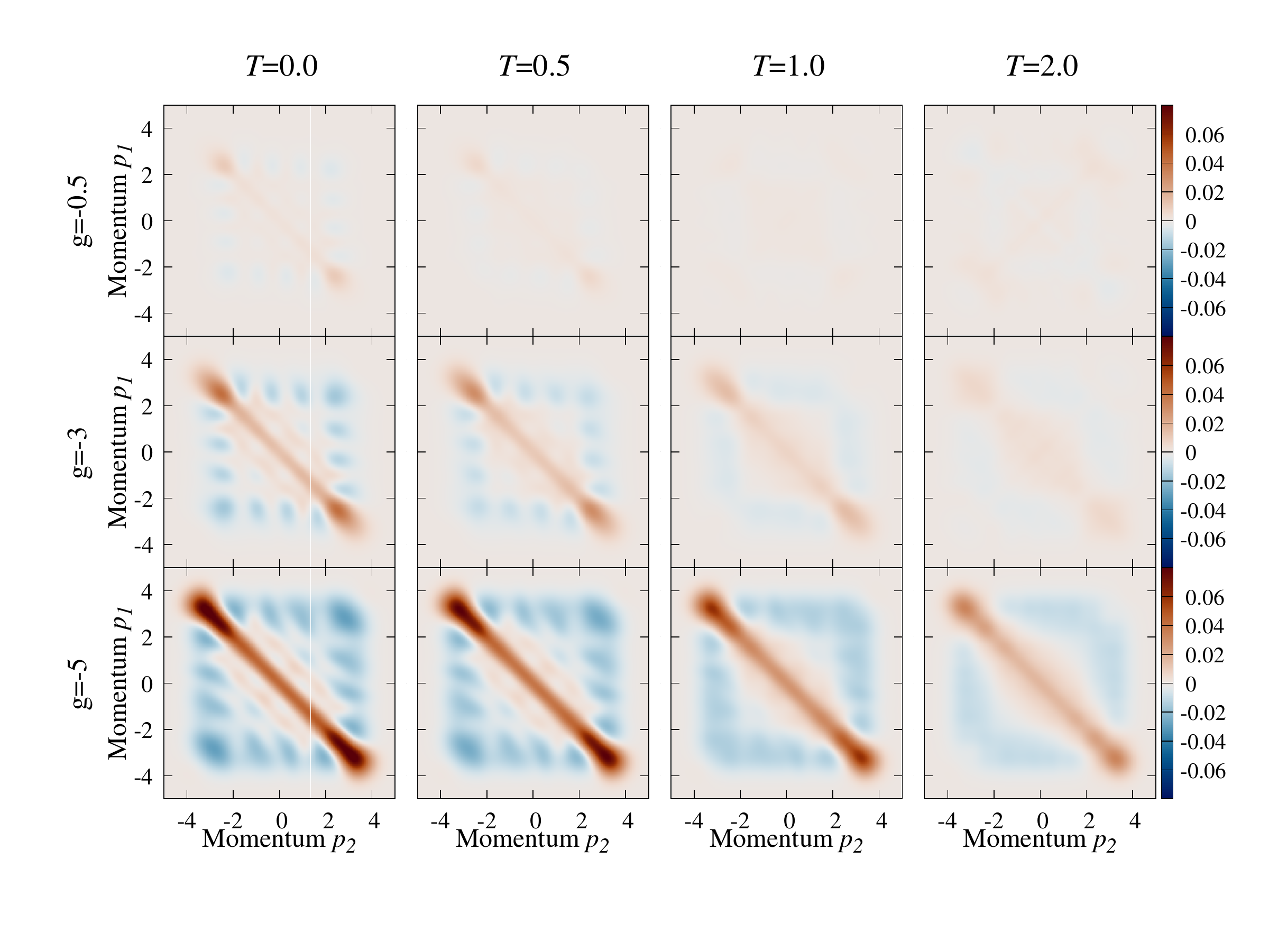}
\caption{
The noise correlation ${\cal G}_T(p_1,p_2)$ calculated for the balanced system of $N=8$ particles. Consecutive rows correspond to different attractions $g$, while columns to different temperatures $T$. Note that along with increasing interactions anti-diagonal correlations are enhanced. Contrary, increasing temperature blurs the noise correlation meaning that the pairing is diminished. All momenta and temperatures are expressed in natural units of the problem, {\it i.e.}, $\sqrt{\hslash m \omega}$ and $\hslash \omega /k_B$, respectively.
\label{Fig2} }
\end{figure}

\section{Balanced systems}\label{sec:temperature}
\begin{figure}
\centering
\includegraphics[width=\linewidth]{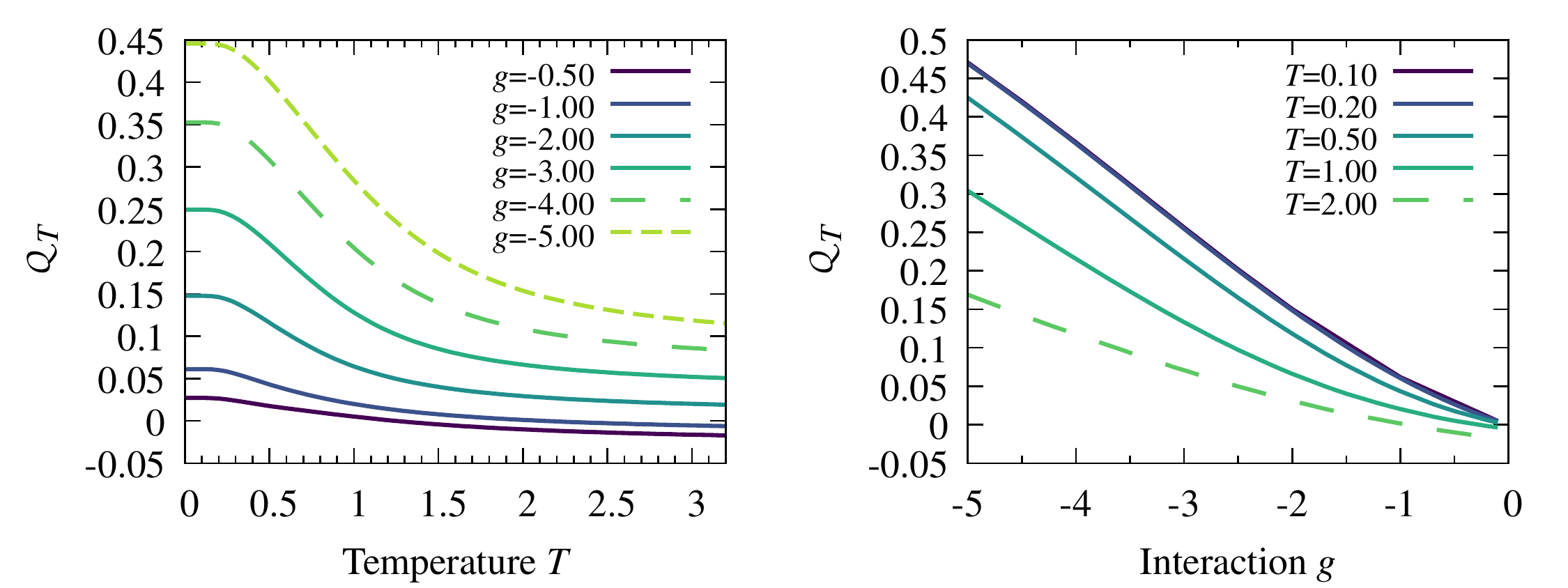}
\caption{The pairing intensity ${\cal Q}_T$ for the balanced system of $N=8$ particles as a function of temperature $T$ (left) and interaction strength $g$ (right). It is clear that decreasing attractions and/or increasing temperature reduce the intensity of pairing. Temperature is expressed in its natural units, $\hslash\omega/k_B$.\label{Fig3}}
\end{figure}

Let us start the analysis from the simplest case of balanced system (polarization $\mathrm{P}=0$) containing in total $N$ particles. In Fig.~\ref{Fig2} we show the dependence of the noise correlation \eqref{ThNoiseCorr} on the interaction strength and temperature for $N=8$ particles. When the system is purely in its many-body ground-state ($T=0$), strong anti-correlations between opposite spin momenta are enhanced with increasing interactions. This signals a strong BCS pairing mechanism studied recently~\cite{2015SowinskiEPL}. It is also clear that this enhancement is reduced by increasing temperature. One can quantify these competing behaviors by integrating the nose correlation close to its anti-diagonal part. For example, it can be done by calculating the pairing intensity ${\cal Q}_T$ expressed as integration with appropriately localized filtering function
\begin{equation} \label{PairingIntensity}
{\cal Q}_T = \int\!dp_1dp_2\, {\cal G}_T(p_1,p_2){\cal F}(p_1+p_2),
\end{equation}
where the function ${\cal F}(p)$ has a gaussian form
\begin{equation}
 {\cal F}(p) = \frac{1}{\sqrt{\pi \kappa}} \exp\left( -\frac{p^2}{2\kappa^2} \right).
\end{equation}
We checked that the final results do not depend on the particular shape of the filter function ${\cal F}$, nor on its width $\kappa$ that should be of reasonable value compared to the resolution of the noise correlation and size of the system. In our calculations we set $\kappa = 0.04\sqrt{\hslash m\omega}$. In Fig.~\ref{Fig3} we display the intensity ${\cal Q}_T$ as a function of temperature for different interaction strengths and as a function of attractions for different temperatures. These results confirm that increasing attractions in the system enhances pairing correlations between opposite spin fermions. At the same time, by increasing temperature, these correlations are suppressed. Note, however, that for some small but finite range of temperatures pairing correlations are almost insensitive to the temperature. This fact is a direct consequence of the finite gap to the excited many-body states mentioned previously. 

\section{Polarized systems}
The situation is more interesting when imbalanced mixtures are considered. In these cases, although inter-component attractions force the system to form correlated pairs, their center-of-mass momentum is not vanishing. In the limit of vanishing temperature ($T\rightarrow 0$) the situation was studied already with all details in \cite{2020PecakPRR}. It was argued that the effect is a direct manifestation of the FFLO mechanism originating in a mismatch of Fermi momenta of interacting components. It was shown that in the case of harmonically trapped particles center-of-mass momentum of the FFLO pair $q_0$ is related to the difference of quasi-Fermi momenta $p_{F\uparrow}-p_{F\downarrow}$, being defined through the Fermi energies $\epsilon_{F\sigma}$ as 
\begin{equation} \label{FermiMom}
p_{F\sigma} = \sqrt{2m\epsilon_{F\sigma}}=\sqrt{2m\hslash\omega(N_{\sigma}-1/2)}.
\end{equation}
In the following, we aim to generalize these findings for finite temperatures.

\begin{figure}[t!]
\includegraphics[width=\linewidth]{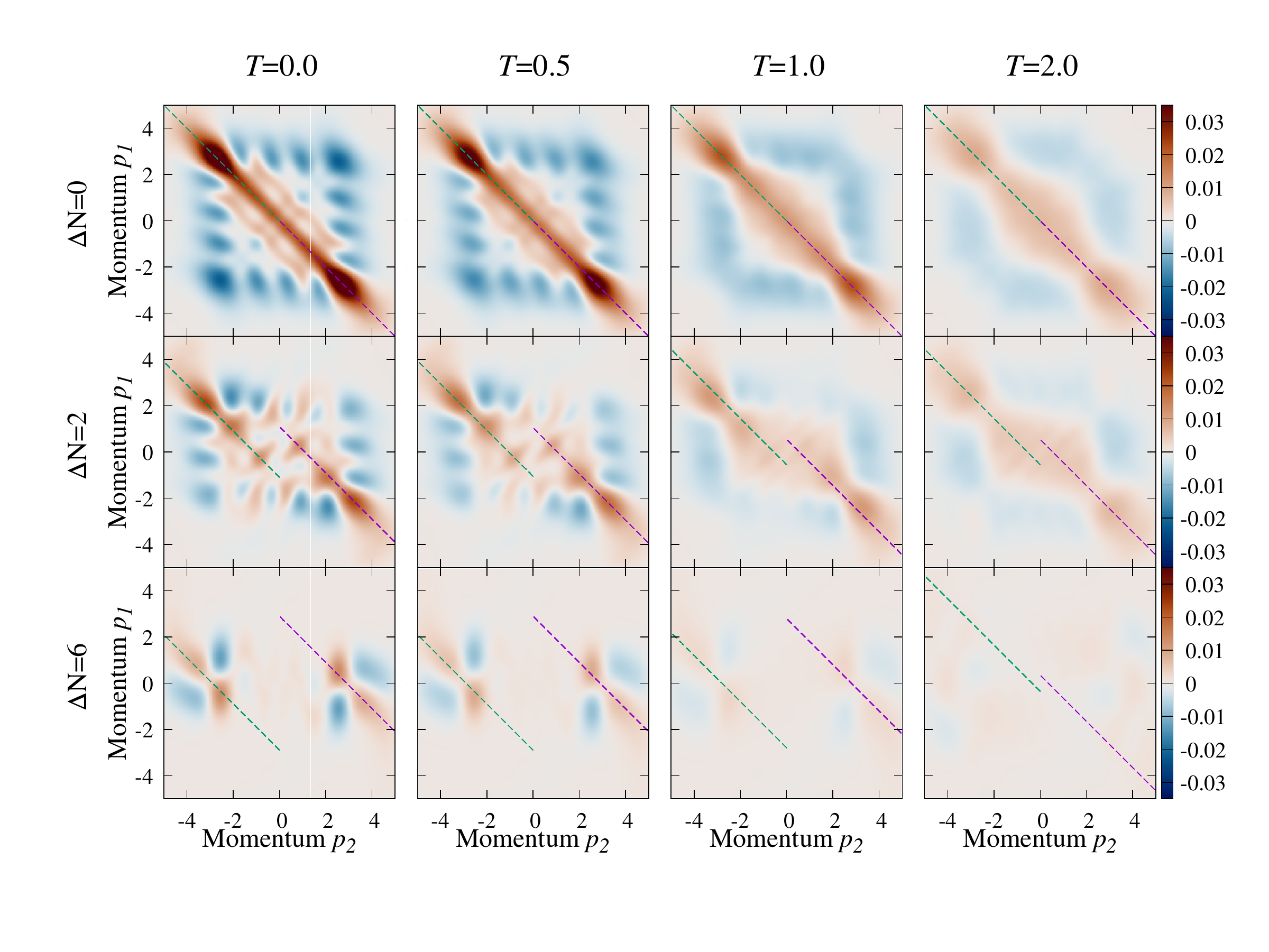}
\caption{
The noise correlation ${\cal G}_T(p_1,p_2)$ calculated for the system of $N=8$ particles at strong attraction $g=-3$. The consecutive rows (columns) correspond to increasing imbalances $\Delta N$ (temperatures $T$). All momenta and temperatures are expressed in natural units of the problem, {\it i.e.}, $\sqrt{\hslash m \omega}$ and $\hslash \omega /k_B$, respectively.
\label{Fig4}
}
\end{figure} 
The analysis starts by displaying noise correlations for different imbalances and temperatures. In Fig.~\ref{Fig4} we show examples for the system containing $N=8$ and for quite strong attraction $g=-3$. The first row (results for the balanced system) corresponds directly to the middle row in Fig.~\ref{Fig2} (note the different range of the scale). The first column, on the other hand, corresponds to the zero-temperature results obtained earlier in \cite{2020PecakPRR}. Indeed, we see, that particle imbalance between components leads directly to a specific splitting of the anti-correlation of opposite-spin momenta and force correlated pairs to have finite center-of-mass-momentum. Along with increasing imbalance, we notice that areas of enhanced pairing keep moving away from the anti-diagonal $p_1+p_2=0$. When the temperature is increased they are not only substantially diminished but also some tiny reversal shift towards the ani-diagonal is visible. To quantify this quite complicated behavior we generalize the pairing intensity \eqref{PairingIntensity} to make it center-of-mass dependent
\begin{equation} \label{PairingIntensity2}
{\cal Q}_T(q) = \int\!dp_1dp_2\, {\cal G}_T(p_1,p_2){\cal F}(p_1+p_2+q).
\end{equation}
The quantity ${\cal Q}_T(q)$ can be understood as a distribution of the net momentum of correlated pairs. For vanishing temperature, the definition \eqref{PairingIntensity2} is identical with the corresponding quantity introduced in \cite{2020PecakPRR}. Such a construction of ${\cal Q}_T(q)$ gives an insight into the most probable center-of-mass momenta of correlated pairs since they are encoded in its maxima. Note that due to the general symmetry of the problem studied, ${\cal Q}_T(q)\equiv{\cal Q}_T(-q)$. Thus in the following, without losing generality, we display results only for non-negative momenta, $q\geq 0$.
\begin{figure}
\centering
\includegraphics[width=\linewidth]{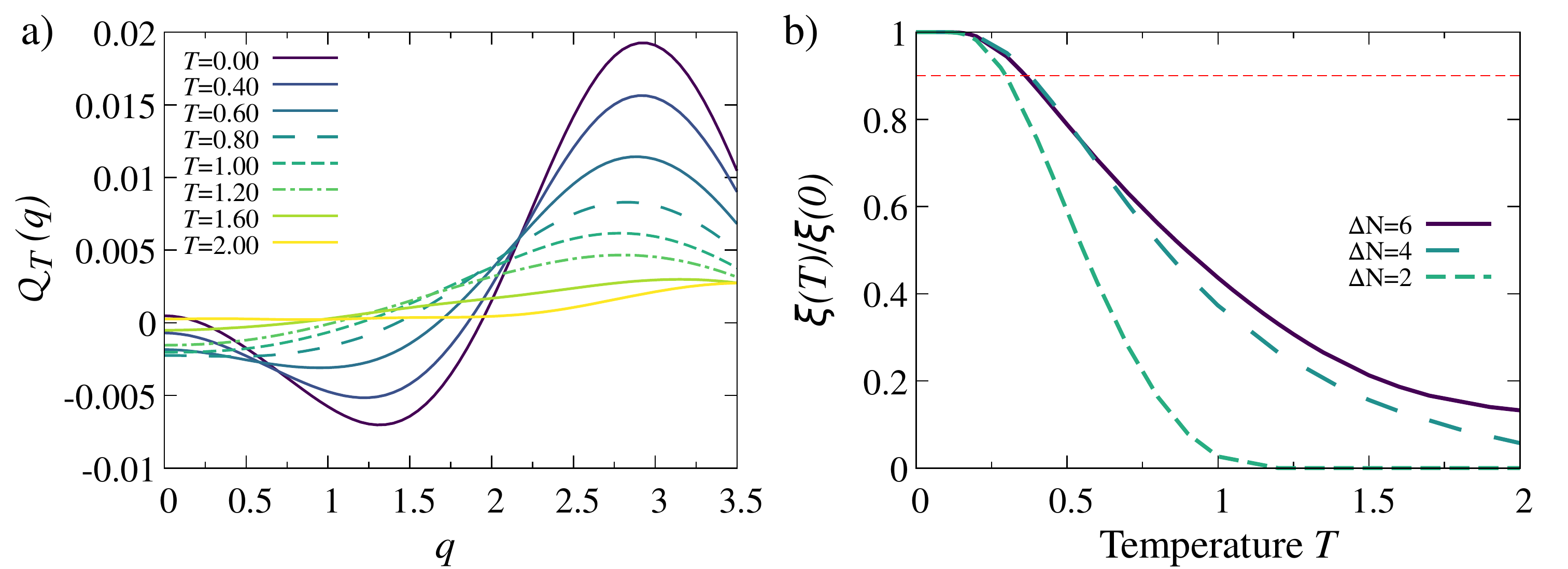}
\caption{Thermal properties of the system with $N=8$ particles and interaction $g=-3$. (a) Distribution of the net momentum of correlated pairs ${\cal Q}_T(q)$ for system with a very large particle imbalance $\Delta N=6$ calculated for different temperatures $T$. Note that increasing temperature reduces the intensity of the maximum and shifts its position towards zero momentum $q=0$. (b) The measure $\xi(T)/\xi(0)$ as a function of temperature for different particle imbalances $\Delta N=2,4,6$. All momenta and temperatures are expressed in natural units of the problem, {\it i.e.}, $\sqrt{\hslash m \omega}$ and $\hslash \omega /k_B$, respectively.
\label{Fig5}
} 
\end{figure}

As a working example, in Fig.~\ref{Fig5}a we show the pairing intensity distribution ${\cal Q}_T(q)$ for the system of $N=8$ particles and the highest imbalance $\Delta N=6$, obtained for different temperatures $T$. For zero temperature a clear maximum at $q_0\approx 2.9\sqrt{m\hslash\omega}$ corresponds directly to the difference of quasi-Fermi momenta predicted by \eqref{FermiMom} and signals strong FFLO pairing with appropriate center-of-mass momentum. 
When the temperature is increased, the amplitude of the maximum at $q_0$ is reduced and its position is shifted toward the second maximum at $q=0$ representing standard BCS pairs which are also present in the system. Since, the maxima for zero and finite momentum coexist for any temperature, it is hard, if possible at all, to witness correlation of a given type in the system independently. As a consequence, for sufficiently large temperature, the distinction between FFLO and BCS pairs is no longer possible. Especially since both pair formation mechanisms are already very weak at this point and it is difficult to refer to any significant existence of correlated pairs. More importantly, as the temperature increases, there is an initial increase in the amplitude of BCS pairs and a simultaneous decrease for FFLO pairs. Finally, for sufficiently large temperatures both maxima melt to a single, very wide plateau meaning that any pairing correlations in the system are not detectable by the noise correlation means. Between these two extreme cases, {\it i.e.} strong FFLO signaled at $T=0$ and lack of any pairing at high $T$ limit, we observe that maximum at $q_0$ drops much faster than the one in the center at $q=0$.
In consequence, for some range of temperatures, their amplitudes are comparable and the system undergoes crossover from unconventional FFLO-like pairing to the standard BCS-like mechanism. A very similar effect has been shown in terms of Bogoliubov-de Gennes formalism~\cite{machida2006generic}, and with Monte Carlo calculations for one-dimensional systems of many fermions confined in optical lattice~\cite{2010WolakPRA}. Our result can be viewed as a few-body precursor of this many-body behavior.

To quantify the mutual relation between the FFLO pairing and the BCS pairing in the system at a given temperature, we introduce the simplest possible measure related to the difference of corresponding maxima
\begin{equation}
\xi(T) = {\cal Q}_T(q_0) - {\cal Q}_T(0).
\end{equation}
We checked that this simple definition gives qualitatively the same predictions as some more sophisticated ones introduced recently \cite{2021DobrzynieckiPRR}.
Since a concrete value of $\xi(T)$ has no direct interpretation, in the following we always normalize it to its zero-temperature value, $\xi(T)/\xi(0)$. With this normalization, value $1$ means that FFLO correlations carried by the finite-momentum pairs are the same as at $T=0$. The temperature at which the quantity starts to drop quickly indicates the moment at which the FFLO phase looses its stability. 

In Fig.~\ref{Fig5}b we show how this dependence looks for systems with $N=8$ particles and different imbalances $\Delta N$. It is clear that for higher imbalances larger temperature is needed to destroy finite-momentum pair correlations. 
Dependence of $\xi(T)/\xi(0)$ on temperature supports our previous findings that the transition from FFLO to BCS pairing induced by temperature has clear features of a crossover rather than a sharp transition as predicted for one-dimensional systems of many particles~\cite{2010WolakPRA}. 
Alike for ${\cal Q}_T$ in balanced systems, $\xi(T)/\xi(0)$ does not change much for very small temperatures. We define the characteristic temperature $T_C$ of the crossover as a point when $\xi(T)/\xi(0)$ drops below 90\% (dashed line in Fig.~\ref{Fig5}b). Such definition captures adequately the moment at which the FFLO-type correlations start decaying and the system is quickly driven towards the BCS regime. We want to emphasize that the chosen threshold 90\% is to some extent arbitrary. Therefore the exact value of $T_C$ will depend on our criterium. However, we checked that the particular definition does not affect the general connection between polarization and temperature which seems to be universal.

From our numerical results, it follows that the characteristic temperature $T_C$ slightly increases with attractions $g$. What is more, for a fixed interaction strength $g$ we found that increasing the imbalance $\Delta N$ makes the tail of $\xi(T)/\xi(0)$ thicker. The reason is simply that the correlations in the system are much more complex than just one of the two types BCS or FFLO, and thus they are beyond noise correlation analysis. Therefore, for sufficiently large imbalances the function $\xi(T)/\xi(0)$ does not drop exactly to zero anymore and after reaching its minimal value starts to grow. However, since the ground state of any imbalanced system is always FFLO-like correlated at $T=0$, our definition of characteristic temperature $T_C$ captures appropriately the physics of the crossover even if pure BCS-like correlation at large temperatures cannot be reached for a given interaction $g$.
 
\begin{figure}
\centering
\includegraphics[width=0.8\linewidth]{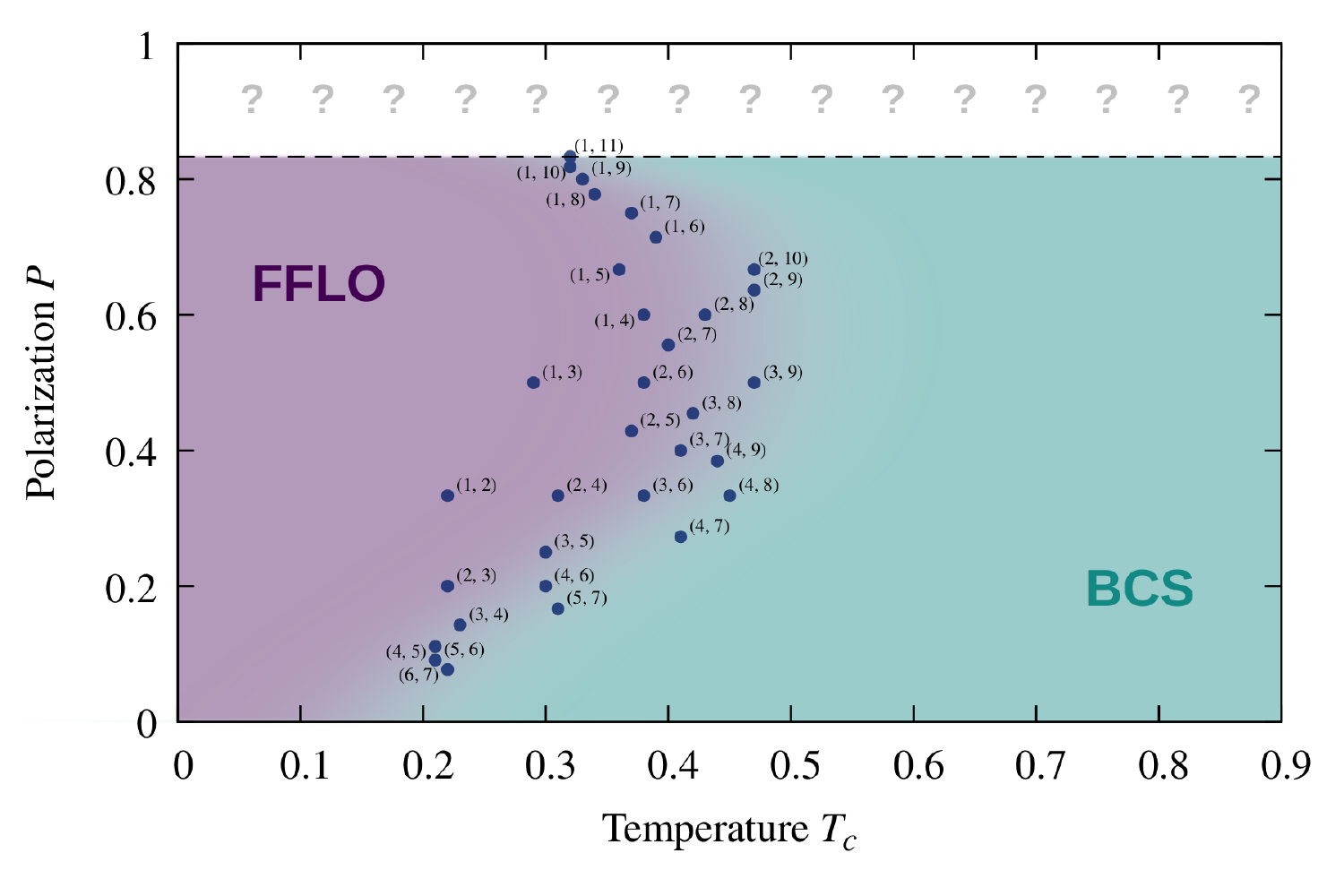}
\caption{
Phase diagram of the attractively interacting few-fermion system at $g=-3$. Different points correspond to different systems with the number of particles $(N_\uparrow,N_\downarrow)$ indicated in parenthesis. Apart from the points ($1$,$N-1$) corresponding to the polaronic-like systems, the other points are clearly arranged along the boundary between the two phases. Critical temperature $T_C$ is expressed in its natural unit, $\hslash \omega /k_B$.\label{Fig6}
}
\end{figure}
\section{Few-body phase diagram}
Having all previous results and observations in hand we can repeat all the reasoning for different numbers of particles and different imbalances. In this way, we are able to form a provisional bridge between few-body systems and the many-body counterparts. We assume that in the limit of large systems appropriate description of the system is served by intensive properties which are independent of the size of the system. In the system studied the polarization \eqref{polarization} plays such a role. Therefore, we suspect that for large enough systems we should approach the many-body limit if properties of the system are presented as functions of polarization $P$ and temperature $T$. It is worth noting that sometimes different many-body properties might saturate unexpectedly quickly (see for example experiments reported in~\cite{2013WenzScience}).

To make the first view on this problem, in Fig.~\ref{Fig6} we display characteristic temperatures of the transition $T_C$ for systems having different polarizations $P$. To make it systematic, we performed calculations for all possible distributions of particles among components with up to $N=12$ particles in total. Different points correspond to systems with different numbers of particles shown in parenthesis. Note, that there are families of points giving the same polarization $P$. We find that points having the largest number of particles align on a quite regular and characteristic curve forming a border between FFLO and BCS pairing. We see that the characteristic temperature $T_C$ grows with the polarization up to $P \approx 0.5$ and then starts to decrease. Due to a numerical limitation of our approach based on the exact diagonalization, we are not able to study systems with $P>0.84$ (separated by a horizontal dashed line). At this point, it should be noted also that some points evidently stand out from the border predicted. These points correspond mainly to polaronic systems containing only one particle in a selected component. Of course, these systems, independently on the polarization $P$ and the total number of particles $N$ cannot be viewed as limiting in the sense that both components have many particles. 
Due to their polaronic nature, one should not suspect that they may display all typical properties of systems hosting the FFLO-like correlations.

The phase diagram obtained is surprisingly very similar to the phase diagram obtained earlier with Monte Carlo calculations for many fermions in a one-dimensional optical lattice~\cite{2010WolakPRA}. Although some quantitative differences are visible, qualitatively the results obtained for these two substantially different systems are compatible. 

\section{Conclusions}\label{sec:conclusions}
In this work, using the exact diagonalization approach, we have examined how finite-temperature effects impact two-body correlations in two-component fermionic systems containing few particles with attractive interactions. We focused on correlations encoded in the so-called noise correlation -- the quantity which in principle is directly accessible in experiments. We have shown that as the temperature increases, the correlations are initially insensitive (due to the energy gap) and from a certain moment they vanish very quickly. 

Importantly, for imbalanced systems hosting the FFLO pairing in their many-body ground state, before pairing correlations are lost, we have observed the transition from the FFLO-like to the BCS-like phase. This transition has features of a crossover rather than a rapid phase transition and can be well characterized by the characteristic temperature $T_C$ interpreted as a temperature when the FFLO-like phase loses its stability. 

By systematic studies of systems having different numbers of particles, we determine the approximate border in the phase diagram between the two phases. Our predictions qualitatively agree with previous results obtained for larger systems confined in periodic potentials and in harmonic traps. As a side result, we showed that polaronic systems containing a single particle in one of the components have typically substantially different properties and should be studied separately. 

\begin{appendix}

\section{The numerical method used}\label{App1}

Our numerical method is based on a numerically exact diagonalization of the many-body Hamiltonian \eqref{eq:hamiltonian} with applied energetic cut-off of the many-body basis~\cite{1998HaugsetPRA,2018PlodzienARX,2019ChrostowskiAPPA,2020RojoMathematics}. First we consider a corresponding single-particle problem of a one-dimensional harmonic oscillator whose eigenstates are well-known and represented by wavefunctions
\begin{equation} \label{singleorbitals}
\phi_k(x) = N_k\,\mathrm{H}_k\left(\sqrt{\frac{m\omega}{\hslash}}x\right)\exp\left(-\frac{m\omega}{2\hslash}x^2\right),
\end{equation} where $\mathrm{H}_k(x)$ is $k$-th Hermite polynomial and $N_k=\left(2^kk!\sqrt{\pi\hslash/m\omega}\right)^{-1/2}$ is a normalization constant. Corresponding eigenergies are equal $E_k=\hslash\omega(k+1/2)$. Having these solutions in hand, for a fixed number of particles $N_\uparrow$ and $N_\downarrow$ we consider all possible Fock states of the form
\begin{equation}
|\mathtt{F}_{\boldsymbol{\alpha}}\rangle = \hat{a}_{i_1}^\dagger\cdots\hat{a}^\dagger_{i_{N_\uparrow}}\hat{b}_{j_1}^\dagger\cdots\hat{b}^\dagger_{j_{N_\downarrow}}|\mathtt{vac}\rangle,
\end{equation}
where $\boldsymbol{\alpha}=(i_1,\ldots,i_{N_\uparrow};j_1,\ldots,j_{N_\uparrow})$ is a set of (sorted in descending order) indices of occupied single-particle orbitals \eqref{singleorbitals} while fermionic operators $\hat{a}_i$ ($\hat{b}_i$) annihilate $\uparrow$-particle ($\downarrow$-particle) in corresponding states $\phi_i(x)$. A non-interacting energy of a Fock state $|\mathtt{F}_{\boldsymbol{\alpha}}\rangle$ is simply a sum of single-particle energies of occupied orbitals
\begin{equation}
\varepsilon_{\boldsymbol{\alpha}}=E_{i_1}+\ldots + E_{i_{N_\uparrow}}+ E_{j_1}+\ldots + E_{j_{N_\downarrow}}.
\end{equation}
By decomposing field operators in the basis of these annihilation operators
\begin{equation} \label{FieldDecomp}
\hat{\Psi}_\uparrow(x)= \sum_k \hat{a}_k \phi_k(x),\qquad \hat{\Psi}_\downarrow(x)= \sum_k \hat{b}_k \phi_k(x)
\end{equation}
one can easily represent the many-body Hamiltonian \eqref{eq:hamiltonian} as a matrix in the basis of Fock states $\{|\mathtt{F}_{\boldsymbol{\alpha}}\rangle\}$. Corresponding matrix elements can be calculated straightforwardly as
\begin{equation}
{\cal H}_{\boldsymbol{\alpha}\boldsymbol{\beta}} = \langle \mathtt{F}_{\boldsymbol{\alpha}}|\hat{\cal H}|\mathtt{F}_{\boldsymbol{\beta}}\rangle = \varepsilon_{\boldsymbol{\alpha}}\delta_{\boldsymbol{\alpha}\boldsymbol{\beta}}+\sum_{ijkl}U_{ijkl}\langle \mathtt{F}_{\boldsymbol{\alpha}}|\hat{b}_i^\dagger\hat{a}_j^\dagger\hat{a}_k\hat{b}_l|\mathtt{F}_{\boldsymbol{\beta}}\rangle,
\end{equation}
where interaction integrals are expressed as
\begin{equation}
U_{ijkl}=g\, \sqrt{\frac{\hslash^3\omega}{m}}\int\!\!dx\, \phi^*_i(x)\phi^*_j(x)\phi_k(x)\phi_l(x).\end{equation}
These integrals can be calculated numerically on a dense spatial grid or analytically by using specific properties of Hermite polynomials~\cite{2020RojoMathematics,rammelmuller2022modular}.

Naturally, in practice, we are not able to consider all possible Fock states since their number grows exponentially with the number of particles and number of single-particle orbitals. Thus some limitation of the Fock basis is required. Since we are interested in the low-lying spectrum of the Hamiltonian \eqref{eq:hamiltonian} and not too far from the non-interacting case ($g=0$) we select from a whole basis the states with the lowest non-interacting energy, {\it i.e.}, only these Fock states whose energies $\varepsilon_{\boldsymbol{\alpha}}$ are not larger than some energetic cut-off $E_C$. Since in the case studied single-particle energies are equally distributed and proportional to their occupation indices, this requirement technically means that we pick to the basis only these Fock states $|\mathtt{F}_{\boldsymbol{\alpha}}\rangle$ for which 
\begin{equation}
i_1+\ldots+i_{N_\uparrow}+j_1+ \ldots+j_{N_\downarrow} \leq E_C - \hslash\omega(N_\uparrow+N_\downarrow)/2.
\end{equation}
A detailed numerical algorithm for quick generation of the Fock basis in this form (also for other confinements) was presented recently~\cite{2019ChrostowskiAPPA}. In our calculations, we select the cut-off $E_C$ to provide well-converged final results. In practice $E_C/\hslash\omega - (N_\uparrow+N_\downarrow)/2$ is never less than 20. 

After all these preparations the resulting matrix ${\cal H}_{\boldsymbol{\alpha}\boldsymbol{\beta}}$ is numerically diagonalized using the Arnoldi alghoritm \cite{ArnoldiBook}. As a result one obtain the lowest eigenenergies ${\cal E}_i$ and corresponding decomposition coefficients $\{f^{(i)}_{\boldsymbol{\alpha}}\}$ representing many-body eigenstates $|\mathtt{i}\rangle$ in the cropped Fock basis:
\begin{equation}
|\mathtt{i}\rangle = \sum_{\boldsymbol{\alpha}} f^{(i)}_{\boldsymbol{\alpha}} |\mathtt{F}_{\boldsymbol{\alpha}}\rangle.
\end{equation}
With this decomposition, one can easily calculate the expectation value of any many-body operator expressed in terms of the field operators $\hat{\Psi}_\sigma(x)$ by decomposing them according to \eqref{FieldDecomp}.
 
\end{appendix}

\section*{Acknowledgements }
T.S. acknowledges fruitful discussions and hospitality at the UPV in Val\`encia. This work was supported by the (Polish) National Science Centre Grants No. 2016/22/E/ST2/00555 (T.S.) and 2021/40/C/ST2/00072 (D.P.).

\section*{Author contributions statement}
D.P., and T.S. equally contributed in all stages of the project. All authors reviewed the manuscript.

\section*{Data availability}
All data generated or analysed during this study are available from the corresponding author on reasonable request.

\section*{Additional information}

\textbf{Competing financial and non-financial interests} All authors declare no competing interests.

\bibliography{_biblio}

\end{document}